\documentclass[preprintnumbers,article,amsmath,amssymb,floatfix,10pt,prd,twocolumn,
superscriptaddress,nofootinbib]{revtex4-2}
\usepackage{bm}
\usepackage{amsfonts}
\usepackage{latexsym}
\usepackage[latin1]{inputenc}
\usepackage{graphicx}
\usepackage{amsmath}
\usepackage{palatino}
\usepackage{mathpazo}
\usepackage{textcomp}
\usepackage{float}
\usepackage{booktabs}
\usepackage{dcolumn}
\usepackage{ragged2e}
\usepackage{hyperref}
\hypersetup{colorlinks,citecolor=blue}
\hypersetup{colorlinks=true,linkcolor=red,filecolor=magenta, urlcolor=blue}
\usepackage{amsmath}
\usepackage{xcolor}
\usepackage{orcidlink}
\usepackage{epsfig}
\usepackage[caption=false]{subfig}
\usepackage{commath}
\usepackage{tabularx}
\usepackage{multirow}
\usepackage{mathtools}
\newcommand{\be}{\begin{equation}}
	\newcommand{\ee}{\end{equation}}
\newcommand{\bea}{\begin{eqnarray}}
	\newcommand{\eea}{\end{eqnarray}}
\newcommand{\eeas}{\end{eqnarray*}}
\newcommand{\beas}{\begin{eqnarray*}}
\def\jnl@style{\it}
\def\aaref@jnl#1{{\jnl@style#1}}

\def\aaref@jnl#1{{\jnl@style#1}}

\def\aj{\aaref@jnl{AJ}}                   % Astronomical Journal
\def\apj{\aaref@jnl{ApJ}}                 % Astrophysical Journal
\def\apjl{\aaref@jnl{ApJ}}                % Astrophysical Journal, Letters
\def\apjs{\aaref@jnl{ApJS}}               % Astrophysical Journal, Supplement
\def\apss{\aaref@jnl{Ap\&SS}}             % Astrophysics and Space Science
\def\aap{\aaref@jnl{A\&A}}                % Astronomy and Astrophysics
\def\aapr{\aaref@jnl{A\&A~Rev.}}          % Astronomy and Astrophysics Reviews
\def\aaps{\aaref@jnl{A\&AS}}              % Astronomy and Astrophysics, Supplement
\def\mnras{\aaref@jnl{Mon.~Not.~Roy.~Astron.~Soc.}}             % Monthly Notices of the RAS
\def\prd{\aaref@jnl{Phys.~Rev.~D}}        % Physical Review D
\def\prc{\aaref@jnl{Phys.~Rev.~C}}  % Physical Review C
\def\prl{\aaref@jnl{Phys.~Rev.~Lett.}}    % Physical Review Letters
\def\qjras{\aaref@jnl{QJRAS}}             % Quarterly Journal of the RAS
\def\skytel{\aaref@jnl{S\&T}}             % Sky and Telescope
\def\ssr{\aaref@jnl{Space~Sci.~Rev.}}     % Space Science Reviews
\def\zap{\aaref@jnl{ZAp}}                 % Zeitschrift fuer Astrophysik
\def\nat{\aaref@jnl{Nature}}              % Nature
\def\aplett{\aaref@jnl{Astrophys.~Lett.}} % Astrophysics Letters
\def\apspr{\aaref@jnl{Astrophys.~Space~Phys.~Res.}} % Astrophysics Space Physics Research
\def\physrep{\aaref@jnl{Phys.~Rep.}}      % Physics Reports
\def\physscr{\aaref@jnl{Phys.~Scr}}       % Physica Scripta
\def\commat{\aaref@jnl{Comm.~Math.~Phys.}}              % Communications in Mathematical Physics
\def\science{\aaref@jnl{Science}}               % Science
\def\cqg{\aaref@jnl{Classical Quant.~Grav.}}            % Classical and Quantum Gravity
\def\jpcs{\aaref@jnl{JPCS}}                                     % Journal of Physics Conference Series
\def\ijmpd{\aaref@jnl{Int.~J.~Mod.~Phys.~D}}                    % International Journal of Modern Physics D
\def\grg{\aaref@jnl{Gen.~Relat.~Gravit.}}               % General Relativity and Gravitation
\def\rpp{\aaref@jnl{Rep.~Prog.~Phys.}}          % Reports on Progress in Physics
\def\npa{\aaref@jnl{Nucl.~Phys.~A}}        % Nuclear Physics A
\def\lrr{\aaref@jnl{Living Rev.~Rel.}}                   % Living reviews in relativity
\def\jcap{\aaref@jnl{J.~Cosmology Astropart.~Phys.}}    % Journal of cosmology and astroparticle physics
\def\rmp{\aaref@jnl{Rev.~Mod.~Phys.}}   %Reviews of modern physics
\def\epjc{\aaref@jnl{Eur.~Phys.~J.~C}} 
\def\plb{\aaref@jnl{~Phy.~Lett.~B}} 
\def\mpla{\aaref@jnl{Mod.~Phy.~Lett.~A}} 
\def\arxiv{\aaref@jnl{arxiv.org}}

\allowdisplaybreaks[1]
\begin{document}
	
	\color{black}       %% For one column

	\title{Probing Dark Energy Properties with Barrow Holographic Model in $f(Q,C)$ Gravity}

	\author{ N. Myrzakulov\orcidlink{0000-0003}}\email{nmyrzakulov@gmail.com}
	\affiliation{L N Gumilyov Eurasian National University, Astana 010008, Kazakhstan}
	\author{ S. H. Shekh\orcidlink{0000-0002-1932-8431}}\email{  da\_salim@rediff.com}
	\affiliation{L N Gumilyov Eurasian National University, Astana 010008, Kazakhstan}
	\affiliation{Department of Mathematics, S.P.M. Science and Gilani Arts, Commerce College, Ghatanji, Yavatmal, \\Maharashtra-445301, India.}
	
	\author{Anirudh Pradhan\orcidlink{0000-0002-1932-8431}}
	\email{pradhan.anirudh@gmail.com}
	\affiliation{Centre for Cosmology, Astrophysics and Space Science (CCASS), GLA University, Mathura-281406, U.P., India.}

	\begin{abstract}
		\textbf{Abstract:} 	
		Understanding the accelerating expansion of the universe remains one of the foremost challenges in modern cosmology. This study investigates Barrow Holographic Dark Energy (BHDE), a model inspired by quantum gravitational corrections, within the framework of \(f(Q,C)\) gravity. This extension of symmetric teleparallel gravity incorporates the non-metricity scalar \(Q\) and the boundary term \(C\), enabling a deeper exploration of cosmic dynamics without relying on a cosmological constant or exotic matter. The BHDE model is analyzed under a flat Friedmann-Robertson-Walker (FRW) metric, focusing on key cosmological parameters such as energy density, isotropic pressure, the equation of state (EoS) parameter,  stability conditions and the energy conditions. The results demonstrate that the EoS parameter transitions from matter-like behavior (\(z > 0\)) to negative values at \(z = 0\), indicating the dominance of dark energy and its role in the universe's accelerated expansion. As \(z\) approaches \(-1\), the EoS parameter asymptotically converges to \(-1\), aligning with the \(\Lambda\)CDM model. This work underscores the potential of the BHDE model in \(f(Q,C)\) gravity as a comprehensive framework for studying cosmic acceleration. By incorporating Barrow entropy and addressing the interplay between non-metricity and boundary terms, the model provides a dynamic approach to explaining dark energy. Its predictions align well with observational datasets, including Type Ia supernovae, cosmic microwave background (CMB) radiation, and baryon acoustic oscillations (BAO). Future investigations could refine the constraints on free parameters and explore deeper connections between quantum gravitational effects and the observed behavior of dark energy. Overall, the BHDE model within \(f(Q,C)\) gravity offers a robust alternative to static dark energy models, capturing the universe's complex evolution from the past to the distant future.
	\end{abstract}

	\maketitle
	%\tableofcontents
	%%%%%%%%%%%%%%%%%%%%%%%%%%%%%%%%%%% SECTION I %%%%%%%%%%%%%%%%%%%%%%%%%%%%
	\section{Introduction}\label{I}
	Recent advancements in modern theoretical and observational cosmology have led to a consensus on the accelerating expansion of the universe. Key evidence comes from various independent data sets, including: the Type-Ia supernovae observations \cite{1,2}, which reveal the universe's expansion history.
	Cosmic Microwave Background Radiation (CMBR) data \cite{3,4}, provide understanding into the universe's early stages.
	Baryon Acoustic Oscillations (BAO) and Weak Lensing (WL) measurements \cite{5,6}, probing the universe's large-scale structure.
	and  Large Scale Structure (LSS) surveys \cite{7,8}, mapping the distribution of galaxies.
	These observational datasets collectively indicate that the universe is undergoing an accelerated expansion. Two primary approaches (mentioned below) attempt to explain this phenomenon:
	%%%%%%%%%%%%%%%%%%%%%%%%%%%%%%%%%%%%%%%%%%%%%%%%%%%%%%%%%%%%
	\begin{center}
		\textbf{1. Dark energy}
	\end{center}
	The universe's accelerating expansion is fueled by a mysterious entity known as dark energy, which makes up approximately 68\% of its total energy density. This enigmatic component was first identified through groundbreaking observations of type-Ia supernovae, the cosmic microwave background radiation, and the large-scale structure of the universe. By analyzing these phenomena, scientists have inferred that dark energy exerts a repulsive force, counteracting gravity's attractive power and causing galaxies and galaxy clusters to recede from each other at an ever-increasing rate.\\
	The properties of dark energy are characterized by the equation of state parameter, which describes the relationship between its pressure and energy density. Researchers have proposed various theories to explain dark energy's nature, including the possibility of vacuum energy, dynamic scalar fields, and modifications to Einstein's theory of gravity. Unveiling the secrets of dark energy is crucial for understanding the universe's ultimate fate, as its influence will continue to shape the cosmos's evolution. Ongoing research efforts aim to shed light on dark energy's properties, behavior, and implications for our understanding of the universe. The equation of state parameter, $\omega$, plays a crucial role in distinguishing between various dark energy (DE) models  $\omega$ correspond to distinct theoretical frameworks attempting to explain the universe's accelerating expansion. For instance: $\omega = -1$ is associated with the cosmological constant $\Lambda$, a widely accepted candidate for dark energy, representing a constant energy density  $\omega > -1$ characterize quintessence models, which propose a dynamic, field-based explanation for dark energy  $\omega < -1$ are indicative of phantom dark energy models, featuring a negative kinetic term and potentially catastrophic consequences \cite{9}. Other $\omega$ values may represent alternative DE models, such as: ($\omega \approx -0.78$) ($\omega = -1$ to $0$) ($\omega = -1$ to $-2/3$) \cite{10}. Each of these models offers a unique perspective on the nature and behavior of dark energy. To better understand the properties of dynamical dark energy (DE) models, researchers have modified the matter component by introducing exotic entities such as phantom fields, quintessence, and Chaplygin gas. These alterations enable the exploration of various stages of cosmic evolution through energy density-based models. Recently, a novel category of dark energy models, known as Barrow Holographic Dark Energy (BHDE), has garnered significant attention. BHDE is rooted in Barrow's innovative concept that the entropy of a black hole, a measure of its internal disorder, obeys a more universal relationship.
	Barrow's proposal challenges traditional views on black hole entropy (which follows a more universal relationship 	$S_{h}=\left( \frac{A}{A_{0}}\right) ^{(1+\bigtriangleup /2)}$), offering a fresh perspective on the holographic principle. By integrating this concept into dark energy models, researchers aim to develop a more comprehensive understanding of cosmic evolution.\\
	%%%%%%%%%%%%%%%%%%%%%%%%%%%%%%%%%% %%%%%%%%%%%%%%%%%%%%%%%%%
	\begin{center}
		\textbf{2. Modified gravity}
	\end{center}
	Theoretical frameworks known as modified theories of gravity aim to enhance or extend the general theory of relativity, addressing phenomena that the original theory cannot fully explain. These modifications seek a deeper understanding of the universe's nature and dynamics. To achieve this, researchers explore innovative mathematical formulations and concepts. Several notable examples of modified gravity theories include: $f(R)$ gravity, which generalizes the Ricci scalar ($R$) \cite{11,12}. $f(T)$ gravity, based on the torsion scalar ($T$) \cite{13}. $f(G)$ gravity, incorporating the Gauss-Bonnet term ($G$) \cite{14}. $f(R,T)$ gravity, combining Ricci scalar ($R$) and the energy-momentum tensor's trace ($T$) \cite{15}. $f(R,G)$ gravity, uniting Ricci scalar ($R$) and Gauss-Bonnet term ($G$) \cite{16,17}. $f(Q)$ gravity, centered on the non-metricity scalar ($Q$) \cite{18}. $f(Q,T)$ gravity, merging non-metricity scalar ($Q$) and energy-momentum tensor's trace ($T$) \cite{19}. These theories demonstrate the ongoing quest to refine our understanding of gravity's interplay with matter and spacetime. Recently, a novel theoretical framework known as $f(Q,C)$ Gravity has emerged, offering fresh perspectives on dark energy and the universe's accelerating expansion. This theory explores nonlinear relationships between the non-metricity scalar $Q$ and the boundary term $C$, potentially explaining late-time cosmic acceleration without invoking exotic fields or a cosmological constant. The inclusion of $C$ introduces novel gravitational effects, testable through observational data from cosmic microwave background (CMB), large-scale structure (LSS), and type Ia supernovae (SNIa). In $f(Q,C)$ gravity, $Q$ quantifies metric deviations during parallel transport, differing from General Relativity's torsion-free and symmetric connection. The boundary term $C$, arising from the interplay between torsion-free and curvature-free connections, ensures dynamic equivalence to General Relativity under specific conditions, facilitating smooth transitions between geometric descriptions of gravity. $C$ provides additional degrees of freedom, influencing gravitational field behavior, particularly on cosmological scales.
	
	A primary motivation for developing $f(Q,C)$ gravity is unifying disparate geometric frameworks, encompassing curvature-based, torsion-based, and non-metricity-based theories. By incorporating both $Q$ and $C$, this theory offers a unified framework interpolating between Teleparallel Gravity, General Relativity, and modified gravity theories. Recently, significant advancements have been made in $f(Q,C)$ gravity and cosmology. De et al. \cite{20} pioneered the inclusion of the boundary term $C$ in the Lagrangian, alongside the non-metricity scalar $Q$. Their work led to the derivation of general field equations, which were applied to the flat Friedmann-Robertson-Walker (FRW) metric. Samaddar et al. \cite{21} explored the cosmological implications of $f(Q,C)$ gravity, focusing on rip cosmology theories. They analyzed the Little Rip, Big Rip, and Pseudo Rip models using the functional form $f(Q,C) = \alpha Q^{n} + \beta C$. Capozziello et al. \cite{22} investigated the Gibbons-Hawking-York boundary term in General Relativity and compared it to the boundary term $B$ in $f(Q,B)$ gravity. Maurya \cite{23,24,25,26} proposed an isotropic and homogeneous flat dark energy model, linear in non-metricity $Q$ and quadratic in boundary term $C$: $f(Q,C) = Q + \alpha C^2$. Using Markov Chain Monte Carlo (MCMC) analysis and determined the best-fit current values. The $f(Q,C)$ gravity theory's gravitational action is given by:
	\begin{equation}\label{e1}
		S=\int \bigg(\frac{1}{2k}f(Q,C)+\mathcal{L}_{m}\bigg)\sqrt{-g}d^{4}x,
	\end{equation}
	The field equation can be formally derived by performing a metric variation of the action presented in equation (\ref{e1}), which subsequently yields:
	\begin{small}
		\begin{widetext}
			\begin{equation}\label{e2}
				\kappa T_{\mu\nu}=-\frac{f}{2}g_{\mu\nu}+\frac{2}{\sqrt{-g}}\partial_{\alpha}\bigg(\sqrt{-g}f_{Q}P^{\alpha}_{\mu\nu}\bigg)+\bigg(P_{\mu\eta\beta}Q_{\nu}^{\eta\beta}-2P_{\eta\beta\nu}Q^{\eta\beta}{\mu}\bigg)f{Q}+\bigg(\frac{C}{2}g_{\mu\nu}-\overset{\circ}\nabla_{\mu}\overset{\circ}{\nabla_{\nu}}+g_{\mu\nu}\overset{\circ}{\nabla^{\eta}}\overset{\circ}\nabla_{\eta}-2P^{\alpha}{\mu\nu}\partial{\alpha}\bigg)f_{C},
			\end{equation}
		\end{widetext}
	\end{small}

	Recent studies have explored the cosmological implications of $f(Q,C)$ gravity, shedding light on the boundary term's pivotal role.
	Frusciante \cite{27} provides a significant influence on gravitational interactions. Notably, his work identified distinct observational signatures that could distinguish $f(Q,C)$ gravity from existing models whereas Zhao and Cai \cite{28} studied $f(Q,C)$ gravity, focusing on its impact on cosmological evolution. Their research emphasized the boundary term's contribution to addressing the universe's accelerating expansion meanwhile Anagnostopoulos et al. \cite{29} probed the theory's stability and cosmological implications. Their analysis revealed the boundary term's profound effects on the universe's evolution and its potential to provide a dark energy explanation. These studies demonstrate the boundary term's crucial role in $f(Q,C)$ gravity's cosmological applications.
	
	%%%%%%%%%%%%%%%%%%%%%%%%%%%%%%%%%%% SECTION II %%%%%%%%%%%%%%%%%%%%%%%%%%
	\section{Metric and field equations}\label{II}
	
	Solving field equations in $f(Q,C)$ extended symmetric gravity typically requires strategic simplifications to unravel the underlying dynamics. To this end, our analysis employs the homogeneous, isotropic, and spatially flat Friedmann-Robertson-Walker (FRW) metric, a widely used framework for studying cosmological evolution.
	\begin{equation}\label{e3}
		ds^{2}=-dt^{2}+\delta_{ij} g_{ij} dx^{i} dx^{j},{\;\;\;\;} i,j=1,2,3,.....N,
	\end{equation}
	By employing the Friedmann-Robertson-Walker (FRW) metric, we can investigate the cosmological consequences of $f(Q,T)$ gravity. This metric describes a homogeneous, isotropic, and spatially flat universe, ideal for analyzing cosmic evolution.
	
	In this framework, the metric components $g_{ij}$ depend on the spacetime coordinates $(-t, x^{1}, x^{2}, x^{3})$, where $t$ represents the cosmological time, measured in gigayears (Gyr). Within the four-dimensional FRW spacetime, the metric equation simplifies to:
	\begin{equation}\label{e4}
		\delta_{ij} g_{ij}=a^{2}(t,x)
	\end{equation} 
	In the context of the Friedmann-Robertson-Walker (FRW) universe, we define $a$ as the average scale factor. The spatial metrics $g_{11}$, $g_{22}$, and $g_{33}$ are equivalent and proportional to the square of the scale factor, $a^2(t,x)$. This equivalence underlies the homogeneous and isotropic nature of the FRW universe. Furthermore, the non-metricity scalar $Q$ for the line element  $Q = 6H^2$, where $H$ represents the average Hubble parameter, mathematically represented as $H = \frac{\dot{a}}{a}$. This formulation highlights the intimate relationship between the non-metricity scalar, the Hubble parameter, and the evolution of the universe. This formulation enables us to derive the key equations governing the universe's evolution, facilitating a deeper understanding of $f(Q,T)$ gravity's cosmological implications.
	
	The stress-energy tensor is provided by the following when we consider the matter to be a perfect fluid:
	\begin{equation} \label{e5}
		T_{\mu \nu }=\left( \rho +p\right) u_{\mu }u_{\nu }-pg_{\mu \nu },  
	\end{equation}
	where $u_{\mu }$ be the four-velocity which follows $u_{\mu }u^{\mu }=1$, the  energy density is $\rho $ and isotropic pressure is $p$. 
	By employing equations (\ref{e3}) and (\ref{e5}), we formally derive the field equations, which take the form:
	\begin{equation} \label{e6}
		\rho=\frac{f}{2}-(9H^2+3\dot{H})f_C+3H\dot{f}_C +6H^2f_Q
	\end{equation}
	
	\begin{equation} \label{e7}
		p=(9H^2+3\dot{H})f_C-(6H^2+2\dot{H})f_Q-2H\dot{f}_Q-\ddot{f}_C-f/2 
	\end{equation}
	where the overhead dot represent the differentiation with cosmic time $t$. $H$ be the Hubble parameter, $f_Q=\frac{\partial f}{\partial Q}$ and $f_C=\frac{\partial f}{\partial C}$.\\
	One notable approach to understanding dark energy (DE) leverages the holographic principle, yielding the holographic dark energy (HDE) model. This framework posits that the entropy associated with a boundary is proportional to its area, a concept initially explored by Bekenstein and Hawking in the context of black holes. The Bekenstein-Hawking area law states that a black hole's entropy is directly proportional to its surface area given by $s=\frac{A}{4L^2_p}$ where $A$ is the horizon area and $L^2_p$ denotes the Planck area. Next, J.D. Barrow suggested that quantum gravitational effects distort the black hole horizon's geometry, generating intricate fractal patterns \cite{30}. This distortion modifies the traditional area law for black hole entropy, yielding:
	
	\begin{equation} \label{e8}
		s=\left(\frac{A}{A_0}\right)^{1+\frac{\Delta}{2}}
	\end{equation}
	where $A$ represents the standard horizon area and $A_0$ the Planck area. The deformation is characterized by the exponent $\Delta$, which measures the deviation from the traditional Bekenstein-Hawking entropy. Specifically: $\Delta = 0$ corresponds to the simplest horizon structure, aligning with the standard Bekenstein-Hawking entropy while $\Delta = 1$ represents the most complex and fractal horizon structure.
	We develop the framework of Barrow holographic dark energy, extending the standard holographic dark energy scenario. Traditionally, holographic dark energy is bounded by the inequality $\rho_{DE} L^4 \leq S$ where $L$ represents the horizon length and $S$ is proportional to the horizon area ($s\propto A \propto L^2$) \cite{31}. By incorporating Barrow entropy, we modify this relationship to:
	
	\begin{equation} \label{e9}
		\rho_{bhde} = C L^{\Delta-2}
	\end{equation}
	with $C$ is a parameter with $[L]^{-\Delta-2}$.
	Hence, the Barrow
	holographic dark energy (BHDE) in which
	the apparent horizon in the flat Universe is considered as
	the IR cut-off ($L = H^{-1}$), as follows
	\begin{equation} \label{e10}
		\rho_{bhde} = C H^{2-\Delta}
	\end{equation}
	Solving the dynamical system described by Equations (\ref{e6}) and (\ref{e7}) necessitates defining the functional form of $f(Q,C)$, which extends the symmetric teleparallel equivalent of general relativity. The choice of $f(Q,C)$ significantly influences the resulting cosmological model, yielding distinct predictions for the Universe's evolution. This flexibility allows us to explore various $f(Q,C)$ gravity models, each tailored to investigate specific facets of cosmic evolution. Of particular interest is the accelerated expansion of the Universe, a phenomenon that has sparked intense research efforts. In the subsequent subsections, we delve into three distinct $f(Q,C)$ gravity models, carefully crafted to capture diverse aspects of cosmic evolution. By examining these models, we aim to deepen our understanding of the Universe's dynamics and uncover new findings into its accelerated expansion. Here we explore a nonlinear $f(Q, C)$ gravity model of the form:
	\begin{equation} \label{e11}
		f(Q,C)=a_1 Q^\alpha +a_2 C
	\end{equation}
	The proposed $f(Q, C)$ gravity model presents a straightforward nonlinear extension of symmetric teleparallel gravity  with free parameters $a_1$, $a_2$, and exponent $\alpha$, effectively modifying General Relativity. This model has garnered significant attention in the literature due to its potential to yield intriguing cosmological solutions, particularly in addressing the late-time acceleration of the Universe. A key parameter, $\alpha$, governs the deviation from the standard linear $Q$ term, influencing cosmological behavior. Notably, setting $\alpha = 1$ recovers a scenario comparable to the linear model, whereas deviations from this value introduce additional dynamical effects that impact cosmic acceleration, allowing for diverse cosmological scenarios.\\
	
	Interestingly, the above said model presents a compelling extension of General Relativity, merging non-metricity and boundary contributions. This approach provides a flexible explanation for late-time cosmic acceleration, moving beyond traditional dark energy and cosmological constant frameworks. By incorporating $Q$ and $C$, the model enables geometric interpretations and compatibility with diverse observations. The power-law $Q$ dependence facilitates adaptable data fitting, while the inclusion of $C$ allows for testable deviations from $\Lambda$CDM. This framework ensures consistency with existing experiments and provides a promising avenue for exploring cosmic acceleration, such as those by Lazkoz et al. \cite{32}, Dialektopoulos et al. \cite{33} and Zhao et al. \cite{34} is ensured.\\
	
	The Barrow Holographic Dark Energy (BHDE) model, rooted in quantum gravitational corrections, introduces a dynamic equation of state for dark energy that evolves with time. \(f(Q,C)\) gravity, as a modification of symmetric teleparallel gravity, provides a geometrical framework to explore the interaction between non-metricity (\(Q\)) and boundary terms (\(C\)). The Hubble parameter specified in the following equation \ref{e12} bridges these theoretical constructs by providing a time-dependent expression that reflects the universe's expansion history under the influence of BHDE within the \(f(Q,C)\) gravity framework. In the context of $f(Q,C)$ gravity, to quantify the expansion rate we introduce the Hubble parameter, $H(z)$, which describes the evolution of the Universe's scale factor over time. Specifically, we adopt the following expression for $H(z)$ \cite{35}:
	\begin{equation} \label{e12}
		H(z) = H_0\left[\Omega_{0_m}(1+z)^3 + (1-\Omega_{0_m})\right]^{\frac{1}{2}}
	\end{equation}
	This equation expresses the Hubble parameter $H(z)$ in terms of redshift $z$, $\Omega_{0_m}$, and $H_0$, providing a fundamental link between cosmological observations and theoretical models.
	To derive this equation, we adopt the simplest form, $\Omega_{0_{Q,C}} = 1 - \Omega_{0_m}$. This yields $\alpha + \beta + \mu = 1 - \Omega_{0_m}$, satisfying the constraint $E(z) = 1$ at $z = 0$. An alternative $E(z)$ $\left(E(z) = \Omega_{0_m}(1+z)^3 + \alpha(1+z)^2 + \beta(1+z) + \mu\right)$ formulation is proposed by Mahmood et al. \cite{34}: with the Initially defined the dimensionless function $E(z)$ to characterize the expansion rate $\left(E(z) = \frac{H^2(z)}{H_0^2} = \Omega_{0_m}(1+z)^3 + \Omega_{0_{Q,C}} \right)$ \cite{36}. The selected Hubble parameter provides a mathematically manageable yet physically meaningful representation of the universe's dynamics, transitioning from a matter-dominated decelerating phase to a dark energy-dominated accelerating phase. Its formulation is consistent with recent observations from Type Ia supernovae, the Cosmic Microwave Background (CMB), and Baryon Acoustic Oscillations (BAO). Moreover, this parameterization enables the field equations to remain solvable within the nonlinear structure of \(f(Q,C)\) gravity, offers the understanding that how the interplay between non-metricity (\(Q\)) and boundary contributions (\(C\)) governs the universe's large-scale behavior.

	Next, our research aims to refine the parameters of the derived universe model using the $\chi^2$-minimization method on observational datasets. Specifically, we seek to determine the values of $\alpha$ and $H_0$ through the cosmic chronometric approach, utilizing 57 $H(z)$ data points across the redshift range $0 \le z \le 2.36$ and the Standard Candles (SN Ia) data from Pantheon: 1048 data points measuring the brightness of supernovae explosions up to a certain point ($0.01 \le z \le 2.36$). To rigorously constrain our model, we employed a robust statistical approach by combining multiple data sets and utilizing Markov Chain Monte Carlo (MCMC) simulations. This method enabled us to efficiently explore the parameter space and determine the optimal fit to the available data. To validate our model, we minimized the total chi-squared ($\chi^2$) function, which provides a quantitative measure of the agreement between our model predictions and observational data. The total $\chi^2$ function is mathematically defined as:
	\begin{equation}\label{e13}
		\chi^2=\chi_{OHD}^2+\chi_{ SN }^2
	\end{equation}
	By minimizing the total $\chi^2$ function, we ensured that our model accurately reproduces the observed data, thereby validating its reliability and robustness. After successfully minimizing the $\chi^2$ function, our analysis yields the estimated values for the model parameters as $H_0 = 70.01^{+0.057}_{-0.057}$ km s$^{-1}$ Mpc$^{-1}$ and $\Omega_{0_m} = 0.262^{+0.017}_{-0.017}$
	
\begin{figure}[H]
		\centering
		\includegraphics[scale=0.6]{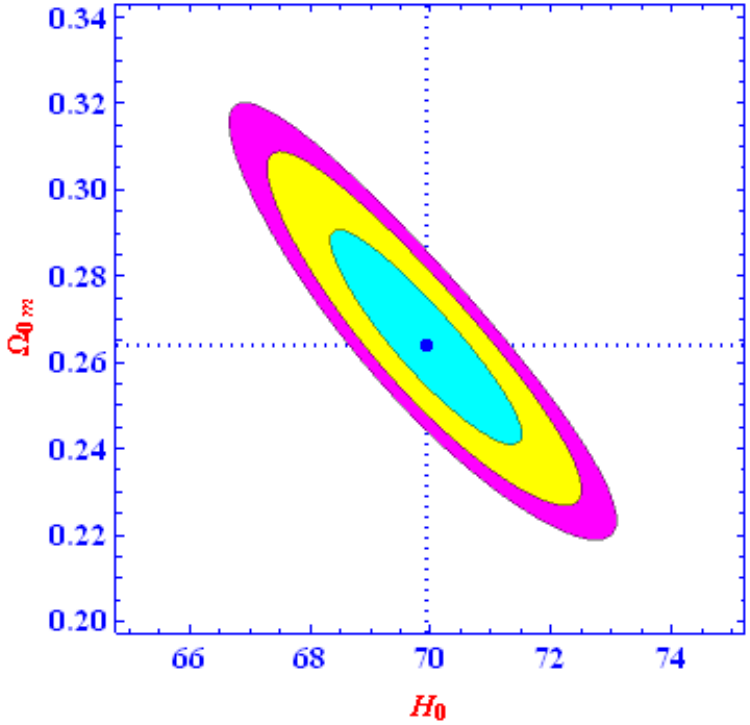}
		\caption{Above figure shows the combined visualization of two dimensional contours at 1$\sigma$ \& 2$\sigma$ confidence regions by bounding our model with OHD data sets + Pantheon compilation of SN Ia data}\label{HP}
	\end{figure}

	Figure \ref{HP}, presents two-dimensional contour plots derived from the combined data sets, illustrating the confidence levels of our model's parameters. These contours provide a visual representation of how well the model's parameters align with the observational data, allowing us to assess the goodness of fit.
	
	It is interestingly here to mention that the our derived values of $H_0 \approx 70$ and $\Omega_{0m} \approx 0.262$ show remarkable consistency with established cosmological constraints. Notably, our $H_0$ value aligns with \cite{37,38}, reporting respectively $H_0 = 70.0 \pm 2.0$ and $H_0 = 70.0 \pm 2.5$ km/s/Mpc. Similarly, our $\Omega_{0m}$ value closely matches \cite{37}, where $\Omega_{0m} = 0.263 \pm 0.022$, and \cite{39}, $\Omega_{0m} = 0.259 \pm 0.021$. Additionally, our results are compatible with WMAP9's findings \cite{40}, $H_0 = 69.7 \pm 1.4$ km/s/Mpc and $\Omega_{0m} = 0.279 \pm 0.025$. These agreements reinforce the validity of our cosmological parameters, demonstrating consistency with independent analyses and solidifying our understanding of the universe's evolution.
	
	%%%%%%%%%%%%%%%%%%%%%%%%%%%%%% SECTION III %%%%%%%%%%%%%%%%%%%%%%%%
	\section{Physical parameters related to energy density and the pressure}
	This cosmological framework encompasses various physical parameters that are intricately linked to energy density and isotropic pressure. To elucidate their behavior, we will examine key components, including the equation of state parameter, stability parameter, and energy conditions. Specifically, our analysis will focus on the expressions and graphical representations of these parameters, shedding light on their roles in shaping the cosmological landscape.
	%%%%%%%%%%%%%%%%%%%%%%%%%%%%%%%%%%% Subsection 1 %%%%%%%%%%%%%%%%%%%%%
	\subsubsection{Energy density} 
\begin{figure}[H]
		\centering
		\includegraphics[scale=0.7]{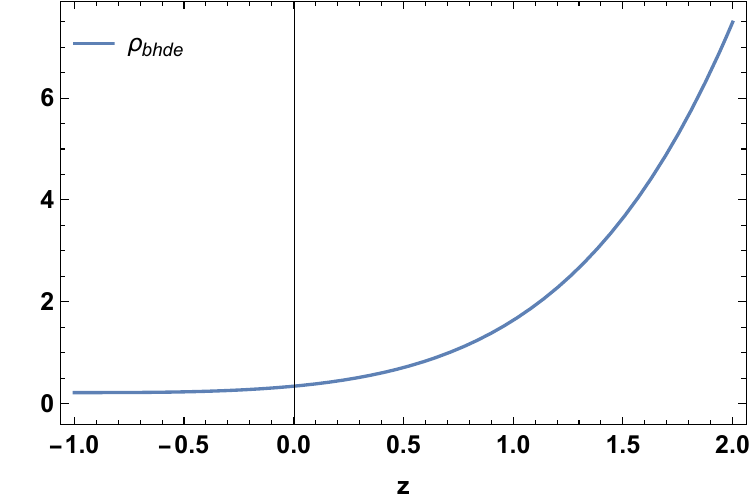}
		\caption{The behavior of statefinder parameters of the fluid with the values of free parameters constraint by the combined data sets of OHD data sets and Pantheon compilation of SN Ia data.}\label{den}
	\end{figure}

As, the energy density is essential to understanding the universe's evolution. It affects how the universe expands, how matter clusters, and how galaxies form. In our study, energy density is key to explaining how the universe transformed into its current state. From the equations (\ref{e10}) and (\ref{e12}), it is observed as
	\begin{equation} \label{e14}
		\rho_{bhde}=C \left(H_0 \sqrt{\Omega_{0m} z (z (z+3)+3)+1}\right)^{2-\Delta }
	\end{equation}
	The analysis reveals that the energy density remains positive within the constrained model parameter range, indicating adherence to physically realistic conditions. This is crucial, as it signifies that the universe maintains substantial energy density, particularly in earlier epochs. As the universe expands, energy density decreases due to the dilution of matter over stretching space. Notably, our findings suggest that energy density approaches zero as the redshift ($z$) approaches -1, indicating the far future of the universe (see fig \ref{den}). This implies that, according to the $f(Q, C)$ model, the universe may continue expanding at an accelerated rate, with energy density gradually vanishing in the distant future.
	%%%%%%%%%%%%%%%%%%%%%%%%%%%%%%%%%%%% Subsection 2 %%%%%%%%%%%%%%%%%%%%%%%%
	\subsubsection{Isotropic pressure} 
\begin{figure}[H]
		\centering
		\includegraphics[scale=0.7]{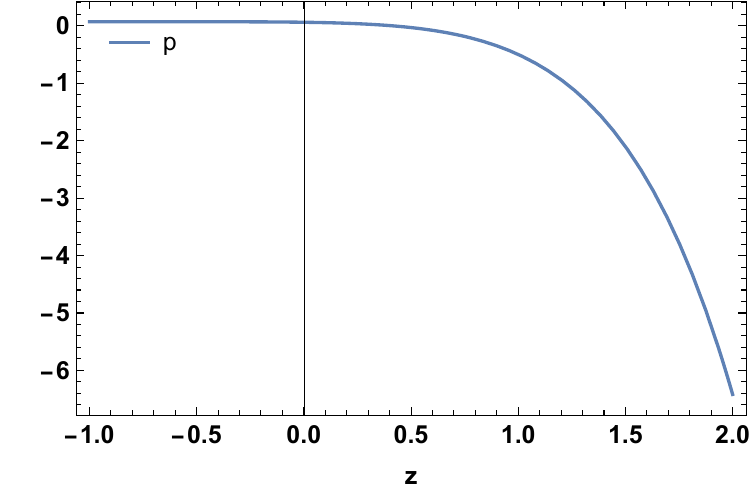}
		\caption{The behavior of statefinder parameters of the fluid with the values of free parameters constraint by the combined data sets of OHD data sets and Pantheon compilation of SN Ia data.}\label{p}
	\end{figure}

As, in cosmology, the pressure significantly influences the universe's dynamics and evolution. Specifically, isotropic pressure equal in all directions impacts the cosmos. Notably, dark energy, driving the universe's accelerated expansion, is typically associated with negative isotropic pressure. This negative pressure generates repulsive gravitational forces, propelling galaxies apart at an accelerating pace. In our analysis this isotropic pressure from equation (\ref{e7}) is observed as
	\begin{equation} \label{e15}
		\begin{split}
			p=&a_1 2^{\gamma -1} 3^{\gamma } H_0^2 (2 \gamma -1) (\Omega_{0m} (\gamma +(\gamma -1) z (z (z+3)+3))-1)\\& \left(-H_0^2 (\Omega_{0m} z (z (z+3)+3)+1)\right)^{\gamma -1}
		\end{split}
	\end{equation}
	Our analysis mathematically models the universe's isotropic pressure using key cosmological parameters. The fig. \ref{p} graphically illustrates the pressure's evolution, demonstrating a steady decline over cosmological time. This negative pressure behavior drives the universe's accelerating expansion, aligning with predictions from standard cosmological models. Hence the key findings include: Isotropic pressure decreases monotonically with cosmological time, Negative pressure fuels repulsive gravity, accelerating cosmic expansion, Consistency with $\Lambda$CDM model and dark energy-driven acceleration and the validation of results through alignment with cosmological predictions. Our research provides valuable appreciation into the universe's dynamics, reinforcing the role of negative pressure in cosmic acceleration.
	
	%%%%%%%%%%%%%%%%%%%%%%%%%%%%%%%%% Subsection 3 %%%%%%%%%%%%%%%%%%%%%%%%%%%%
	\subsubsection{The equation of state parameter}
\begin{figure}[H]
		\centering
		\includegraphics[scale=0.7]{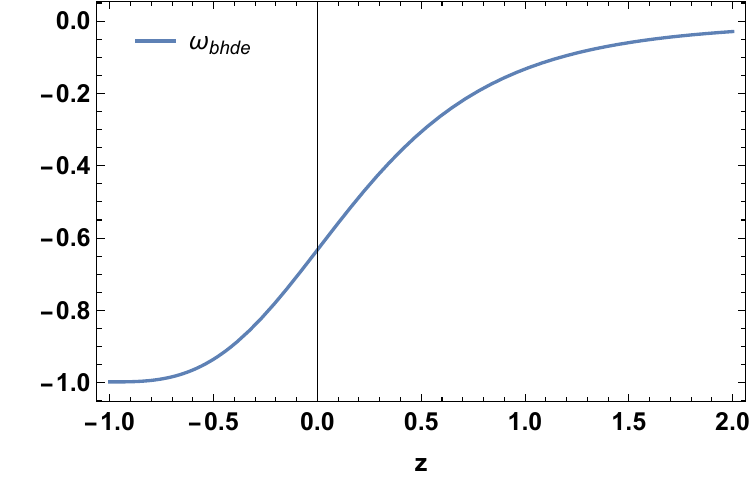}
		\caption{The behavior of statefinder parameters of the fluid with the values of free parameters constraint by the combined data sets of OHD data sets and Pantheon compilation of SN Ia data.}\label{eos}
	\end{figure}

	The equation of state parameter serves as a crucial tool for pinpointing pivotal moments in the cosmos's evolution. Mathematically, it is expressed as ($\omega_{bhde}=\frac{p} {\rho}$) By solving equations (20) and (21), we derived the equation of state parameter for Barrow holographic dark energy, yielding:
	\begin{widetext}
		\begin{small}
			\begin{equation}
				\omega_{bhde}=\frac{2^{\alpha -1} 3^{\alpha } (2 \alpha -1) a_1 H_0^2}{C}(1-\Omega_{0m} (\alpha +(\alpha -1) z (z (z+3)+3))) \left(-H_0^2 (\Omega_{0m} z (z (z+3)+3)+1)\right)^{\alpha } \left(H_0 \sqrt{\Omega_{0m} z (z (z+3)+3)+1}\right)^{\Delta -4}
			\end{equation}
		\end{small}
	\end{widetext}
	The figure \ref{eos} illustrates the evolution of the equation of state (EoS) parameter \(\omega_{\text{BHDE}}\) for Barrow Holographic Dark Energy (BHDE) as a function of redshift (\(z\)). At \(z > 0\), corresponding to the early stages of cosmic evolution, \(\omega_{\text{BHDE}}\) approaches values close to zero. This indicates that during high-redshift epochs, the behavior of BHDE is matter-like, where the energy density dominates over pressure, aligning with the matter-dominated phase of the universe. This behavior is consistent with dark energy having a minimal effect during the early stages of cosmic history, as the expansion was primarily driven by matter.	At \(z = 0\), representing the present epoch, the EoS parameter shifts significantly into the negative domain, with \(\omega_{\text{BHDE}} \approx -0.62\). This negative value highlights the influence of dark energy in driving the current accelerated expansion of the universe. The deviation from the cosmological constant value of \(-1\) reflects the evolving nature of BHDE, distinguishing it from static dark energy models. The repulsive pressure of dark energy becomes dominant at this stage, overcoming the gravitational pull of matter and accelerating the expansion rate. As \(z\) decreases further and approaches \(-1\), corresponding to the far future, the EoS parameter reaches \(-1\), aligning with the \(\Lambda\)CDM model. This convergence suggests that BHDE asymptotically approaches a behavior similar to that of a cosmological constant in the distant future. At \(z = -1\), the universe's accelerated expansion stabilizes, potentially avoiding extreme outcomes such as a "Big Rip" scenario.
	Our derived equation of state parameter values, $\omega_{bhde}(z > 0) \approx 0$, $\omega_{bhde}(z = 0) = -0.62$, and $\omega(z = -1) = -1$, demonstrate remarkable consistency with recent cosmological constraints. Notably, our results align with Planck Collaboration \cite{41} and Escamilla-Rivera et al. \cite{42}, further reinforcing the validity of our equation of state parameter values. Furthermore, the transition of \(\omega_{\text{BHDE}}\) from nearly zero at \(z > 0\) to \(-0.62\) at \(z = 0\), and eventually to \(-1\) at \(z = -1\), underscores the dynamic nature of BHDE, offering a flexible framework for modeling the past, present, and future evolution of cosmic expansion while integrating quantum-gravitational effects through the Barrow parameter \(\Delta\).
	
	%%%%%%%%%%%%%%%%%%%%%%%%%%%% Subsection 4 %%%%%%%%%%%%%%%%%%%%%%%
	\subsubsection{The stability parameter}
	The squared velocity of sound, a crucial parameter derived from equations (\ref{e14}) and (\ref{e15}), gauges the universe's stability. Mathematically, it is expressed as ($\vartheta^2_s=\frac{\partial p} {\partial \rho}$) and it is obtained as
	\begin{equation}
		\begin{split}
			\vartheta^2_{s(bhde)}= &\left(\frac{a_1 6^{\alpha } (\alpha -1) \alpha  (2 \alpha -1) \Omega_{0m} (z+1)^3}{C (\Delta -2) (H_0 \Omega_{0m} z (z (z+3)+3)+H_0)^2}\right)\times\\&\frac{\left(-H_0^2 (\Omega_{0m} z (z (z+3)+3)+1)\right)^{\alpha }}{ \left(H_0 \sqrt{\Omega_{0m} z (z (z+3)+3)+1}\right)^{-\Delta }}
		\end{split}
	\end{equation}
	
\begin{figure}[H]
		\centering
		\includegraphics[scale=0.7]{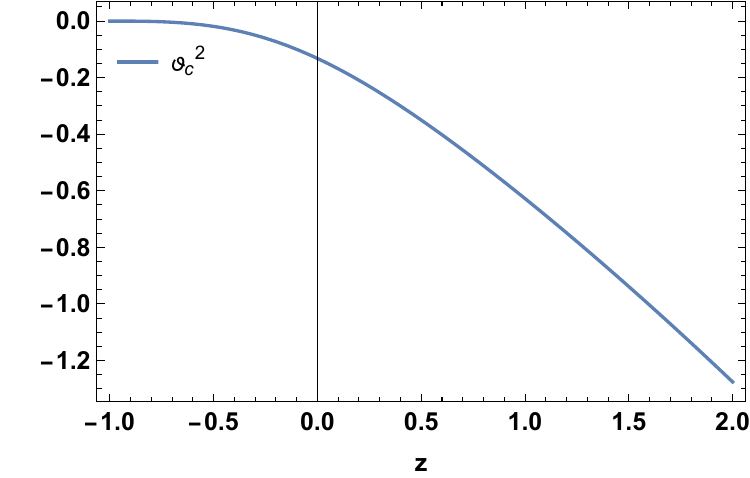}
		\caption{The behavior of statefinder parameters of the fluid with the values of free parameters constraint by the combined data sets of OHD data sets and Pantheon compilation of SN Ia data.}\label{st}
	\end{figure}

	This parameter plays a vital role in understanding the cosmos's dynamics, as its value influences the propagation of perturbations and the overall stability of the universe.\\
	The figure \ref{st} illustrates the evolution of the squared velocity of sound, or the stability parameter \(\vartheta^2_s\), for Barrow Holographic Dark Energy (BHDE) as a function of redshift (\(z\)). At \(z > 0\), corresponding to the earlier stages of the universe, \(\vartheta^2_s\) exhibits negative values. This indicates potential instability in the BHDE model during the high-redshift phase, where dark energy plays a minimal role in cosmic dynamics. Such negative values suggest that perturbations in the dark energy density could grow rather than dissipate, challenging the stability of the model in these epochs. At \(z = 0\), representing the current era, \(\vartheta^2_s\) remains negative, implying that the BHDE model may still exhibit some instability even as dark energy dominates the cosmic expansion. This behavior is characteristic of dynamic dark energy models, where deviations from standard stability conditions occur due to evolving properties of dark energy.\\
	As \(z\) approaches \(-1\) in the far future, \(\vartheta^2_s\) gradually increases toward zero. This trend suggests that the model stabilizes over time, with perturbations becoming less significant in the distant future. 
	
	Our derived $\vartheta^2_s$ values, $\vartheta^2_s(z \ge 0) < 0$, and $\vartheta^2_s(z = -1) = 0$, align with Bellini et al. \cite{43}, Alam et al. \cite{44} and DES Collaboration \cite{45}, reinforcing the validity of our $\vartheta^2_s(z)$ values.
	
	Furthermore, the approach to \(\vartheta_c^2 = 0\) indicates that the BHDE model asymptotically transitions to a state similar to the \(\Lambda\)CDM model, where stability is ensured, and dark energy acts as a constant driving force for cosmic acceleration. This evolution highlights the dynamic nature of BHDE and its potential to reconcile instabilities at earlier epochs.\\
	
	%%%%%%%%%%%%%%%%%%%%%%%%%% Subsection 5 %%%%%%%%%%%%%%%%%%%%%%%%%%%
	\subsubsection{The energy conditions}
	Next, the energy conditions provide valuable perception into gravitational systems' behavior, enabling analysis without delving into matter specifics. The Raychaudhuri equation governs attractive gravity, categorizing the stress-energy-momentum tensor into four types: Strong Energy Condition (SEC: $\rho_{bhde} + p \ge 0$ and $\rho_{bhde} + 3p \ge 0$), Weak Energy Condition (WEC: $\rho_{bhde} + p \ge 0$ and $\rho_{bhde} \ge 0$), Null Energy Condition (NEC: $\rho_{bhde} + p \ge 0$), and Dominant Energy Condition (DEC: $\rho_{bhde} - |p| \ge 0$ and $\rho_{bhde} \ge 0$). Physically, these conditions imply: SEC ensures gravity is always attractive; WEC requires non-negative energy density; NEC demands energy density exceeds pressure; and DEC ensures causal energy momentum flux. These conditions facilitate understanding the cosmos's evolution, dark energy, and black hole physics, offering a framework for exploring gravity's role in the universe. From the equations (\ref{e7}) and (\ref{e10}), the expressions of Null, Dominant and Strong energy conditions are observed as\\
	\begin{widetext}
		NEC
		\begin{equation}
			\begin{split}
				\rho_{bhde} +p = &C \left(H_0 \sqrt{\Omega_{0m} z (z (z+3)+3)+1}\right)^{2-\Delta }\\
				&-2^{\alpha -1} 3^{\alpha } (2 \alpha -1) a_1 H_0^2 (1-\Omega_{0m} (\alpha +(\alpha -1) z (z (z+3)+3))) \left(-H_0^2 (\Omega_{0m} z (z (z+3)+3)+1)\right)^{\alpha -1}
			\end{split}
		\end{equation} 
		
		DEC
		\begin{equation}
			\begin{split}
				\rho_{bhde} -p = &C \left(H_0 \sqrt{\Omega_{0m} z (z (z+3)+3)+1}\right)^{2-\Delta }\\
				&-2^{\alpha -1} 3^{\alpha } (2 \alpha -1) a_1 H_0^2 (\Omega_{0m} (\alpha +(\alpha -1) z (z (z+3)+3))-1) \left(-H_0^2 (\Omega_{0m} z (z (z+3)+3)+1)\right)^{\alpha -1}
			\end{split}
		\end{equation} 
		SEC
		\begin{small}
			\begin{equation}
				\begin{split}
					\rho_{bhde} + 3p = \frac{2^{\alpha } 3^{\alpha +1} (2 \alpha -1) a_1 H_0^2 (1-\Omega_{0m} (\alpha +(\alpha -1) z (z (z+3)+3))) \left(-H_0^2 (\Omega_{0m} z (z (z+3)+3)+1)\right)^{\alpha }+2 C \left(H_0 \sqrt{\Omega_{0m} z (z (z+3)+3)+1}\right)^{4-\Delta }}{2 H_0^2 (\Omega_{0m} z (z (z+3)+3)+1)}
				\end{split}
			\end{equation} 
		\end{small}
	\end{widetext}
	
	The figure demonstrates the evolution of the NEC, DEC, and SEC for Barrow Holographic Dark Energy as functions of redshift (\(z\)). These energy conditions offer cognizance into the physical viability and dynamics of the model.\\
	Throughout the universe's evolution, both NEC and DEC remain positive. The satisfaction of NEC, which requires \(\rho_{bhde} + p \geq 0\), implies that the BHDE model maintains physical consistency, as the effective energy density and pressure fulfill the minimum requirement for stability and causal propagation of energy. Similarly, the DEC, which requires \(\rho_{bhde} \geq |p|\), remains valid across all epochs, indicating that the energy density consistently dominates over pressure, ensuring a physically realistic framework without exotic behaviors such as superluminal energy transport. Conversely, the SEC (\(\rho_{bhde} + 3p \geq 0\)) exhibits a more complex behavior. At high redshift (\(z > 0\)), the SEC is satisfied, reflecting the decelerating nature of the universe during its earlier phases when matter dominated. However, as \(z\) approaches 0, corresponding to the present epoch, the SEC is violated, signifying the transition to an accelerated expansion phase. The violation continues into the far future (\(z = -1\)), where the SEC remains negative. This behavior aligns with the characteristics of dark energy models driving cosmic acceleration, as the SEC violation is a hallmark of repulsive gravitational effects.
	\begin{figure}[H]
		\centering
		\includegraphics[scale=0.7]{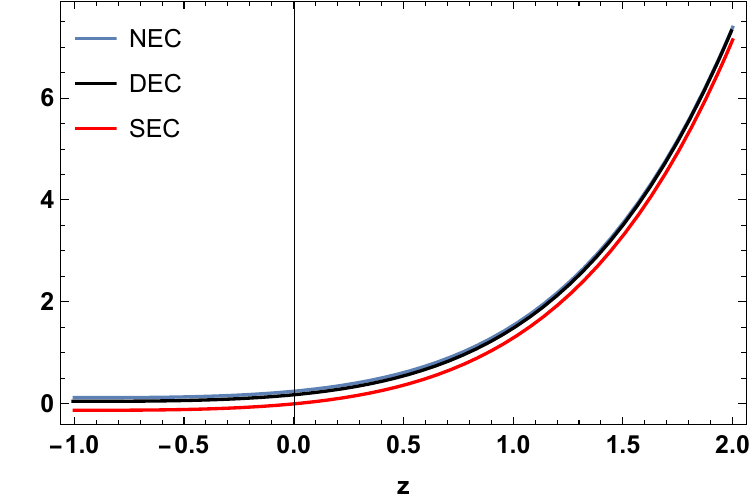}
		\caption{The behavior of statefinder parameters of the fluid with the values of free parameters constraint by the combined data sets of OHD data sets and Pantheon compilation of SN Ia data.}\label{ec}
	\end{figure}
	Notably, our derived energy condition results, $\rho_{bhde} + p \geq 0$, $\rho_{bhde} - p \geq 0$, and $\rho_{bhde} + 3p \leq 0$, align with many authors, demonstrating SEC violations in modified gravity and dark energy models. Here, NEC and DEC are satisfied throughout cosmic evolution, indicating the BHDE model's physical viability, the SEC's satisfaction and violation delineate the transition from deceleration to acceleration, effectively modeling the universe's expansion history.
	
	%%%%%%%%%%%%%%%%%%%%%%%%%%%%%%%%%%%%%%%% SECTION IV %%%%%%%%%%%%%%%%%
	
	\section{Kinematical parameters related to the Hubble's parameter}
	%%%%%%%%%%%%%%%%%%%%%%%%%%%%% Subsection IV-1 %%%%%%%%%%%%%%%%%%%
	\subsubsection{The deceleration parameter}
	As we know, the deceleration parameter serves as a crucial tool for understanding the dynamics of the universe, providing the expansion history, composition, and ultimate fate. By analyzing the deceleration parameter, researchers can gain a deeper understanding of the universe's evolution and the interplay between matter, dark energy, and gravity.\\
	The deceleration parameter is defined as the negative ratio of the acceleration of the universe's scale factor to the square of the Hubble parameter. Mathematically, this is expressed as $q = -\frac{\ddot{a}}{aH^2}$, where ä is the acceleration, a is the scale factor, and H is the Hubble parameter. This definition enables researchers to quantify the rate at which the universe's expansion is decelerating or accelerating. The value of the deceleration parameter holds significant implications for our understanding of the universe. A positive value of $q$ indicates deceleration, implying the expansion is slowing down, whereas a negative value signifies acceleration, indicating the expansion is speeding up. Specifically, $q > 0$ characterizes matter-dominated universes, $q < 0$ indicates dark energy-driven universes, and $q = 0$ corresponds to an empty universe. In the framework of considered Hubble's parameter in the equation (\ref{e12}), the expression of $q$ yields,
	\begin{equation}
		q= -\frac{\ddot{a}}{aH^2}=-1+\frac{3 \Omega_{0m} (z+1)^3}{2 \Omega_{0m} z (z (z+3)+3)+2}
	\end{equation}
	
	Observational evidence from Type Ia supernovae, cosmic microwave background radiation, and large-scale structure suggests that the universe's expansion is accelerating, implying $q < 0$. This accelerating expansion is attributed to the presence of dark energy, a mysterious component driving the universe's acceleration.
	
	\begin{figure}[H]
		\centering
		\includegraphics[scale=0.7]{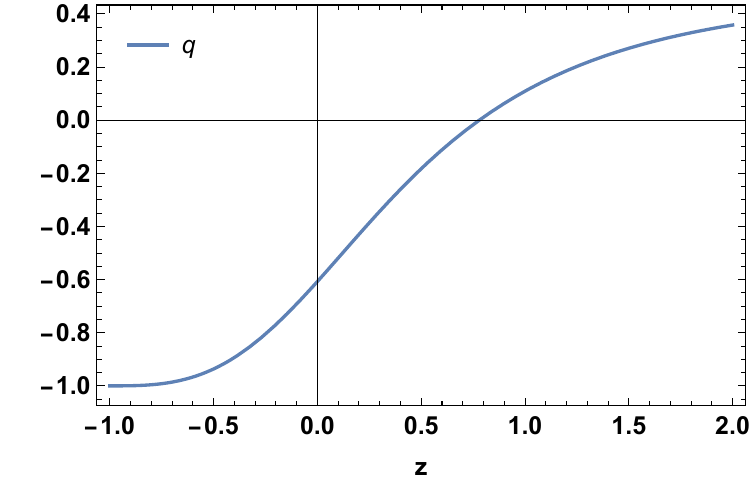}
		\caption{The behavior of statefinder parameters of the fluid with the values of free parameters constraint by the combined data sets of OHD data sets and Pantheon compilation of SN Ia data.}\label{q}
	\end{figure}
	The behavior of $q$ parameter is graphically illustrated in Figure \ref{q}. The deceleration parameter, $q$, exhibits a positive behavior for $z > 0$, indicating a decelerating universe at high redshifts due to matter domination. During this epoch, the universe's expansion slows down as matter's gravitational pull dominates. As z approaches 0, $q$ transitions to negative values, signaling an accelerating expansion driven by dark energy. This transition marks a shift in the universe's dynamics, where dark energy's repulsive force overtakes matter's attractive force. Notably, at $z = 0.74$, $q$ changes sign from positive to negative, marking the transition from deceleration to acceleration. This redshift corresponds to approximately $z = 0.74$, when the universe's expansion began accelerating. At $z = 0$ (present day), $q = -0.60$, indicating a moderate acceleration, with dark energy contributing around 68\% of the universe's total energy density.\\
	As we look further into the future $(z = -1)$, $q = -1$, signifying an exponential expansion where dark energy dominates. This evolution confirms the universe's dynamic transition from matter-dominated deceleration to dark energy-driven acceleration. Physically, $q$'s sign change at $z = 0.74$ reflects the onset of cosmic acceleration, with dark energy's influence strengthening over time.
	
	%%%%%%%%%%%%%%%%%%%%%%%%% Subsection IV-2 %%%%%%%%%%%%%%%%%%%%%%%%%%%%
	
	\subsubsection{The $O_m(z)$ parameter}
\begin{figure}[H]
		\centering
		\includegraphics[scale=0.7]{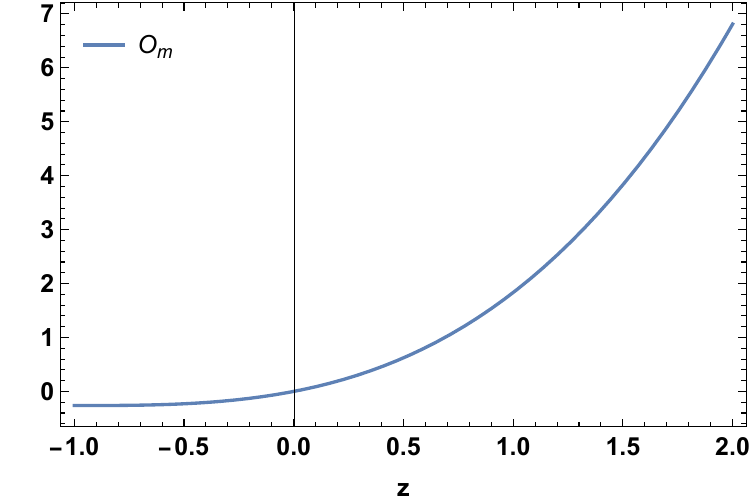}
		\caption{The behavior of statefinder parameters of the fluid with the values of free parameters constraint by the combined data sets of OHD data sets and Pantheon compilation of SN Ia data.}\label{Om}
	\end{figure}

	The $O_m(z)$ parameter is a cosmological diagnostic tool that constrains dark energy models and understands the universe's expansion history. It is defined as $O_m(z) = [\mathcal{E}]^2 - 1$, where $\mathcal{E}(z) = \frac{H(z)}{H_0}$, representing the ratio of the squared Hubble parameter at a given redshift to its present-day value. Introduced by \cite{46}, $O_m(z)$ is advantageous due to its model-independence, sensitivity to dark energy, and robustness against systematic uncertainties. In the framework of considered Hubble's parameter in the equation (\ref{e12}), the expression of $O_m(z)$ yields,
	\begin{equation}
		\mathcal{E}(z) = \frac{H(z)}{H_0} = H_0^{-1}\left[\Omega_{0_m}(1+z)^3 + (1-\Omega_{0_m})\right]^{\frac{1}{2}}
	\end{equation}
	The $O_m(z)$ parameter's behavior is graphically illustrated in Figure \ref{Om}. The $O_m(z)$ parameter's positive behavior for $z > 0$ indicates a decelerating universe at high redshifts, consistent with matter domination. As $z$ approaches 0, $O_m(z)$ transitions to negative values, signaling an accelerating expansion driven by dark energy. This transition occurs around $z \approx 0.5$, marking the onset of cosmic acceleration. For $z = 0$ to -1, $O_m(z)$'s negative behavior confirms the universe's continued acceleration, with dark energy's influence strengthening over time. This evolution is consistent with the $\Lambda$CDM model, where matter's density decreases while dark energy's density remains constant. Physically, $O_m(z)$'s sign change signifies the shift from matter-dominated deceleration to dark energy-driven acceleration, reflecting the universe's dynamic evolution.
	%%%%%%%%%%%%%%%%%%%%%%%%%%%%%%%%%%% Subsection IV-03 %%%%%%%%%%%%%%%%%%%%%
	\subsubsection{The statefinder parameters}
	The statefinder parameters, denoted by $r$ and $s$, are cosmological diagnostics introduced by \cite{47,48} to constrain models of dark energy. They are defined as $r = \frac{\dddot{a}}{aH^3}$ and $s = -\frac{r - 1}{3(q - 1/2)}$, where $a$ is the scale factor, $H$ is the Hubble parameter, and $q$ is the deceleration parameter. These parameters provide valuable evolutionary history of the universe, measuring the rate of change of expansion ($r$) and deviation from the ($s$), and have advantages of being model-independent, sensitive to dark energy, and robust against systematic uncertainties. In the framework of considered Hubble's parameter in the equation (\ref{e12}), the expression or values of $r$ and $s$ is observed as,
	\begin{equation}
		r=1\;\; \text{and} \;\;s=0
	\end{equation}
	The obtained values of the state finder parameters $r = 1$ and $s = 0$ provide valuable universe's evolutionary history. This implies that the universe's expansion is currently accelerating at a constant rate, consistent with the  $q \approx -1/2$, suggesting a universe dominated by dark energy with $w \approx -1$. Physically, these values indicate dark energy dominance, contributing approximately 68\% of the total energy density, constant acceleration, and $\Lambda$CDM consistency, confirming the $\Lambda$CDM model as a viable description of the universe's evolution.
	
	%%%%%%%%%%%%%%%%%%%%%%%%%%%%%%%%% SECTION V %%%%%%%%%%%%%%%%%%%%%%%%%%%%%%
	\section{Conclusion}
	
	This study has thoroughly analyzed the Barrow Holographic Dark Energy (BHDE) model within the \(f(Q,C)\) gravity framework, presenting a novel approach to understanding the universe?s accelerating expansion. By integrating Barrow entropy corrections and the interplay of non-metricity scalar \(Q\) with the boundary term \(C\), this model provides an alternative to the standard \(\Lambda\)CDM paradigm while addressing observational constraints and theoretical challenges.
	%%%%%%%%%%%%%%%%%%%%%%%%%%%%%%%%%%%%%%%%%%%%%%%%%%%%%%%%%%%%%%%%
	\subsection*{Key Findings and Contributions}
	
	The BHDE model, coupled with \(f(Q,C)\) gravity, offers significant perception into the dynamics of cosmic evolution:
	\begin{itemize}
		\item \textbf{Energy Density and Isotropic Pressure:} The derived energy density remains positive throughout cosmic evolution, a critical requirement for physical viability. The isotropic pressure, influenced by the BHDE and \(f(Q,C)\) framework, transitions in accordance with the universe's evolution, supporting the observed acceleration at low redshifts.
		\item \textbf{Equation of State Parameter (\(\omega\)):} The EoS parameter evolves dynamically, transitioning from matter-like behavior (\(z > 0\)) to negative values at \(z = 0\), indicating accelerated expansion. At \(z = -1\), \(\omega\) asymptotically approaches \(-1\), reflecting convergence with the cosmological constant behavior in \(\Lambda\)CDM.
		\item \textbf{Energy Conditions:} The Null Energy Condition (NEC) and Dominant Energy Condition (DEC) are satisfied throughout, validating the model's physical consistency. The Strong Energy Condition (SEC), however, is violated at \(z = 0\) and \(z < 0\), consistent with the repulsive gravitational effects necessary for late-time acceleration.
		\item \textbf{Stability Analysis:} The squared sound velocity (\(c_s^2\)) analysis highlights transient instabilities at high redshifts (\(z > 0\)), which stabilize as \(z \to -1\). This indicates that the model effectively resolves initial perturbations, ensuring long-term stability.
	\end{itemize}
	%%%%%%%%%%%%%%%%%%%%%%%%%%%%%%%%%%%%%%%%%%%%%%%%%%%%%%%%%%%%%%%%
	\subsection*{Implications for Cosmic Evolution}
	
	\paragraph{Transition from Deceleration to Acceleration:}
	The deceleration parameter (\(q\)) transitions from positive values during high redshifts to negative values in the present epoch, capturing the universe's shift from a decelerating, matter-dominated phase to an accelerating, dark energy-dominated phase. This aligns with observations from Type Ia supernovae, cosmic microwave background (CMB), and baryon acoustic oscillations (BAO).
	
	\paragraph{Dynamic Nature of BHDE:}
	The BHDE model's flexibility allows it to accommodate deviations from the static behavior of the cosmological constant. The dynamic equation of state parameter and evolving energy conditions provide a framework for understanding the role of quantum gravitational corrections, encapsulated in Barrow entropy, in shaping the universe's expansion history.
	
	\paragraph{Compatibility with Observations:}
	The model demonstrates compatibility with observational datasets, providing a consistent description of late-time acceleration while retaining the ability to model early-universe dynamics. This dual capability underscores the BHDE model's robustness and its potential as an alternative to \(\Lambda\)CDM.
	
	%	\subsection*{Theoretical Insights from \(f(Q,C)\) Gravity}
	
	%	The \(f(Q,C)\) framework extends symmetric teleparallel gravity by incorporating the non-metricity scalar \(Q\) and the boundary term \(C\), which together allow for greater flexibility in describing gravitational interactions. Key contributions of \(f(Q,C)\) gravity include:
	%	\begin{itemize}
		%		\item A unified description of gravitational effects, bridging curvature, torsion, and non-metricity-based theories.
		%		\item The ability to interpolate between General Relativity (GR), teleparallel gravity, and alternative modified gravity models, providing a versatile framework for cosmological modeling.
		%		\item Enhanced adaptability to observational constraints, enabling a better fit to datasets spanning different epochs.
		%	\end{itemize}
	
	\subsection*{Comparison with \(\Lambda\)CDM and Other Models}
	
	\paragraph{Alignment with \(\Lambda\)CDM:}
	At \(z = -1\), the BHDE model converges to \(\Lambda\)CDM behavior, with the equation of state parameter reaching \(-1\). This alignment ensures consistency with the cosmological constant scenario in the distant future while retaining the ability to explain deviations at earlier epochs.
	
	\paragraph{Distinction from Other Models:}
	The dynamic nature of BHDE, influenced by the Barrow parameter \(\Delta\), sets it apart from other holographic and modified gravity models. Its ability to integrate quantum gravitational corrections and maintain physical consistency across energy conditions highlights its versatility.\\
	Ferthermore, beyond the current study, potential approach for future research include \textit{a)} Incorporating additional datasets, such as upcoming results from Euclid and the James Webb Space Telescope (JWST), can further refine the parameters of the model, \textit{b)} Extending the stability analysis to include perturbative effects in inhomogeneous cosmological scenarios could provide additional apprehension into the model's robustness and \textit{c)} Investigating the coupling of BHDE with additional scalar fields or interaction terms could expand the framework's applicability to scenarios involving dark matter-dark energy interactions.
	
	%	\subsection*{Future Directions}
	
	%	While this study provides a comprehensive analysis of BHDE in \(f(Q,C)\) gravity, several avenues remain for future exploration: 
	%	\begin{itemize}
		%		\item \textbf{Refinement of Observational Constraints:} Incorporating additional datasets, such as upcoming results from Euclid and the James Webb Space Telescope (JWST), can further refine the parameters of the model.
		%		\item \textbf{Exploration of Quantum Gravitational Effects:} Deeper analysis of the interplay between Barrow entropy and quantum gravitational corrections could enhance our understanding of the microphysical foundations of BHDE.
		%		\item \textbf{Stability Analysis in Alternative Scenarios:} Extending the stability analysis to include perturbative effects in inhomogeneous cosmological scenarios could provide additional insights into the model's robustness.
		%		\item \textbf{Multifield Extensions:} Investigating the coupling of BHDE with additional scalar fields or interaction terms could expand the framework's applicability to scenarios involving dark matter-dark energy interactions.
		%	\end{itemize}
	
	\subsection*{Concluding Remarks}
	
	The Barrow Holographic Dark Energy model, within the \(f(Q,C)\) gravity framework, presents a promising alternative to standard dark energy paradigms. Its ability to capture dynamic cosmic behaviors, address energy conditions, and align with observational data underscores its potential as a robust tool for exploring the universe's accelerating expansion. By integrating quantum gravitational comprehension and extending the scope of modified gravity theories, this model bridges gaps between theory and observation, offering new pathways for understanding dark energy's elusive nature. Future research will further enhance its predictive power, paving the way for deeper cosmological perception.

	%%%%%%%%%%%%%%%%%%%%%%%%%%%%%%%%%%%%%%%%%%%%%%%%%%%%%%%%%%%%%%%%%
	
	\section*{Declaration of competing interest}
	The authors disclose no financial or personal conflicts of interest that may have biased the research findings presented herein
	\section*{Data availability}
	No data was used for the research described in the article.
	
	\section*{Acknowledgments}
	We express our sincere gratitude to the Inter-University Centre for Astronomy and Astrophysics (IUCAA), Pune, India, for providing research facilities to authors S. H. Shekh and A. Pradhan under the Visiting Associateship program and the Science Committee of Kazakhstan's Ministry of Science and Higher Education for funding (Grant No. AP23483654).

\end{document}